# Flexible all-PM NALM Yb:fiber laser design for frequency comb applications: operation regimes and their noise properties

ALINE S. MAYER,[1,4] WILFRID GROSINGER,[1] JAKOB FELLINGER,[1] GEORG WINKLER,[1] LUKAS W. PERNER,[1] STEFAN DROSTE,[2] SARPER H. SALMAN,[2] CHEN LI,[2] CHRISTOPH M. HEYL,[2,3] INGMAR HARTL,[3] AND OLIVER H. HECKL[1,5]

[1]*Christian Doppler Laboratory for Mid-IR Spectroscopy and Semiconductor Optics, Faculty Center for Nano Structure Research, Faculty of Physics, University of Vienna, 1090 Vienna, Austria*
[2]*Deutsches Elektronen-Synchrotron DESY, 22607 Hamburg, Germany*
[3]*Helmholtz-Institute Jena, 07743 Jena, Germany*
[4]*aline.mayer@univie.ac.at*
[5]*oliver.heckl@univie.ac.at*

**Abstract:** We present a flexible all-polarization-maintaining (PM) mode-locked ytterbium (Yb):fiber laser based on a nonlinear amplifying loop mirror (NALM). In addition to providing detailed design considerations, we discuss the different operation regimes accessible by this versatile laser architecture and experimentally analyze five representative mode-locking states. These five states were obtained in a 78-MHz configuration at different intracavity group delay dispersion (GDD) values ranging from anomalous (-0.035 ps$^2$) to normal (+0.015 ps$^2$). We put a particular focus on the characterization of the intensity noise as well as the free-running linewidth of the carrier-envelope-offset (CEO) frequency as a function of the different operation regimes. We observe that operation points far from the spontaneous emission peak of Yb (~1030 nm) and close to zero intracavity dispersion can be found, where the influence of pump noise is strongly suppressed. For such an operation point, we show that a CEO linewidth of less than 10-kHz at 1 s integration can be obtained without any active stabilization.

## 1. Introduction

Over the past few years, the performance of low-noise mode-locked fiber lasers has tremendously progressed [1,2]. The growing number of applications of mode-locked lasers in various fields of science is fueling the development of affordable sources that are robust, reliable and can be operated hands-off by people who are not necessarily specialized in laser science. Fiber lasers offer many advantages in that respect: for common gain media such as ytterbium (Yb)- and erbium (Er)-doped glass fibers, pump light from low-cost 976/980-nm semiconductor diode lasers can be delivered to the gain fiber through all-fiber components, thus avoiding pump misalignment. The large surface-to-volume ratio of optical fibers eases thermal management and even long oscillator cavities can be spooled to fit within very small footprints. However, the robustness and reliability is also affected by the type of mode-locking mechanism used. In particular, lasers based on non-polarization maintaining (non-PM) fibers tend to be very sensitive to environmental perturbations such as temperature changes, humidity or mechanical stress. Nevertheless, mode-locking based on nonlinear polarization evolution (NPE) [3–5] in such fibers has been one of the most common techniques to achieve pulsed operation, as it is simple to implement with off-the-shelf optical components and allows for a large flexibility in cavity design.

Since the artificial saturable absorption effect of NPE only occurs if the polarization is capable of rotating freely within the fiber, this mode-locking scheme is not compatible with the use of polarization-maintaining (PM) fibers. PM fibers however significantly increase the robustness against environmental effects. Real saturable absorbers can instead be used to mode-lock lasers based on PM fibers: examples include semiconductor absorber mirrors (SESAMs) [6,7], graphene [8,9], carbon nanotubes (CNT) [10–12], or topological insulators [13,14]. However, the complexity of designing and fabricating such saturable absorber structures and their tendency to degrade over time often hinders their wide-spread implementation.

Another way of achieving mode-locking in polarization-maintaining (PM) fibers consists of using a nonlinear optical loop mirror (NOLM) [15] or nonlinear amplifying loop mirror (NALM) [16]: these devices are based on the superposition of two waves counter-propagating in a fiber loop that can be formed by connecting the output ports of a conventional fiber coupler. The transmission/reflection of the loop mirror depends on the difference in nonlinear phase shift $\Delta\varphi_{nl}$ between the two counter-propagating pulses. Since the nonlinear phase shift is intensity-dependent, these devices can act as saturable absorbers and thus support mode-locking if the NOLM/NALM is implemented such that the total laser cavity round-trip losses drop as soon as $\Delta\varphi_{nl}$ deviates from zero. In a NOLM, the difference in phase shift can be generated by simply having a non-50:50 splitting ratio at its entrance. In a NALM, $\Delta\varphi_{nl}$ is additionally determined by an amplifying fiber asymmetrically placed inside the loop. The ends of the PM fibers used for the NOLM/NALM can be connected to form a so-called "figure-8 laser" [17–20]. Although mode-locking is possible, achieving self-starting operation is difficult in a standard figure-8 configuration [19,21], since small deviations of $\Delta\varphi_{nl}$ from zero only lead to small changes in the cavity losses. To solve this challenge, one can introduce a non-reciprocal phase bias [22] in order to have a steeper slope of the loss curve around $\Delta\varphi_{nl}=0$. The phase bias can be generated by a combination of wave plates and Faraday rotators used in transmission or arranged in the form of a compact reflection module [23]. Recently, it has been shown that an intrinsic phase bias can also be generated by using a 3x3 fiber coupler (instead of 2x2) [24]. Although using a phase bias solves the self-starting issue, the figure-8 configuration still suffers from other drawbacks: it is difficult to scale towards higher repetition rates (i.e. shorter fiber lengths) and lacks an end-mirror that would allow for repetition rate stabilization with a simple piezo actuator.

Thus, a modification of this scheme has recently been proposed [25,26], where the NALM is implemented in reflection (i.e. containing a straight arm with a mirror at its end). Due to the straight arm, the scheme has been named "figure-9 laser" and has been demonstrated both with a NALM-entrance featuring a fixed splitting ratio [25,27,28], as well as using a variable splitting configuration [26]. The concept was applied to lasers based on Yb-, Er-, as well as thulium-holmium (Tm/Ho)-doped gain fibers [26]. Repetition rates as high as 700 MHz have been demonstrated in an Yb-fiber laser using this type of architecture [29], although a non-PM gain fiber had to be used in this case since a PM-fiber with sufficient doping was not commercially available. Nevertheless, due to the strong bending of the short piece of gain fiber, the authors report almost PM-behavior and thus robust operation. The all-PM figure-9 principle has also shown to be compatible with power scaling: using a large mode area (LMA) Yb:fiber, an average output power of 2 W at 72-MHz repetition rate has been reported [30].

Due to their robustness as well as their reliable and reproducible self-starting behavior, all-PM NALM lasers have rapidly turned into ideal candidates for a large variety of applications where "hands-off" mode-locked lasers are desirable [31]. In particular, they have raised considerable interest as sources of low-noise optical frequency combs: full frequency comb stabilization (i.e. stabilization of both the carrier-envelope offset frequency $f_{CEO}$ and comb repetition rate $f_{rep}$) has been shown in Er-based configurations [25,32], whereas $f_{CEO}$-stabilization has also been reported in an Yb-based system [33,34]. Furthermore, even all-PM NALM-based dual-comb configurations have been demonstrated, where the two pulse trains are generated within the same laser cavity [35–37]. In our own work on an all-PM NALM dual-comb system [37], we made the interesting observation that these two pulse trains could actually each run in different intracavity dispersion regimes simultaneously.

These promising results prove that the concept of all-PM NALM is very versatile. However, they also reveal the current lack of a systematic overview of the large parameter space that can be accessed already with such a single-comb system.

Here, we provide practical guidelines for the design and operation of an all-PM NALM mode-locked laser based on Yb in particular and will discuss how to leverage the degrees of freedom to find low-noise operation points.

The paper is organized as follows: In section 2, we discuss detailed cavity design considerations that simultaneously allow for reliability and flexibility. In section 3, we present in-situ cavity dispersion measurements to map the operation regions of the laser. In section 4, we show five representative mode-locking (ML) states obtained in our 78-MHz Yb:fiber configuration, before comparing their free-running comb noise properties (amplitude noise, timing jitter and linewidth of the offset frequency) in section 5 and 6. We will show that operation points far from the spontaneous emission peak of Yb situated at ~1030 nm and close to zero intracavity dispersion can be found, where the influence of pump noise is strongly suppressed. In such a regime, we show that the free-running CEO linewidth can be lower than 10 kHz at 1 s integration time. The sub-10-kHz linewidth presented here rivals the performance reported for non-monolithic solid-state lasers with comparable repetition rates [38]; sub-1-kHz free-running CEO linewidths have so far only been reported for an ultra-low noise fully monolithic 1-GHz solid-state laser [39].

## 2. Laser setup and mode-locking principle

### 2.1 Laser setup

**Table 1. List of components used for the 78-MHz all-PM NALM Yb:fiber laser**

| Component | Description |
| --- | --- |
| Gain fiber | 41.5 cm PM Yb-doped single-mode silica fiber (CorActive Yb-401-PM) |
| Pump diode | 976 nm, Fiber-Bragg-Grating (FBG)-stabilized, max. 900 mW (Thorlabs BL976-PAG900) |
| Pump/Temperature controller | Thorlabs CLD1015, temperature set to 25.0°C |
| Pump isolator | Operation range: 976 ± 10 nm, PM 980 fiber, fast axis blocked, max. average optical power 2W (AFR HPMI-976-02-N-B-QF-1-C) |
| Wavelength division multiplexer (WDM) | Pass band 980 ± 10 nm, reflection band 1060 ± 40 nm, max. optical power (cw): 3W (AFR PMFWDM-9806-N-B-Q) |
| Single mode PM fiber | Corning PANDA PM980 |
| Beam combiner with integrated collimating lens | Specified operation range: 1030 ± 10 nm, polarization aligned to slow axis at both ports, specified output beam waist position (from lens tip): 100-150 mm, max. optical power (cw): 3W (AFR Semi-PBCC-03-09-N-B-Q) |
| Faraday rotator | Rotation @ 1030 nm: 45° ± 0.5° rotator (EOT HP-05-R-1030) |
| Waveplates | $\lambda/2$: Thorlabs WPH05M-1030, $\lambda/4$: Thorlabs WPH05M-1030 |
| Polarizing beamsplitter cube (PBC) | Operation range: 900-1300 nm, Thorlabs PBS103 |
| Gratings | 1000 lines/mm, angle of incidence (Littrow) 31.3°, diffraction efficiency for s-and p >94%, (LightSmyth T-1000-1040-3212-94) |
| End mirror | Silver mirror (Thorlabs PF10-03-P01) |

The laser used for our experimental study is depicted schematically in Fig. 1(a) and a detailed list of its components can be found in Table 1. The cavity consists of a PM NALM fiber portion comprising the Yb-doped gain fiber and a free-space linear section consisting of polarization optics and a grating compressor. The geometrical length of the free-space part from the tip of the collimator lens to the silver end mirror amounts to ~21 cm. The ends of the PM single-mode fiber forming the NALM-loop are twisted by 90° and directly attached to a birefringent beam combiner, which also contains an integrated collimation lens (see Table 1 for details). Light exiting the collimator passes through a 45°-Faraday rotator, a quarter-wave plate (fast axis angle $\theta_q$) and a half-wave plate ($\theta_h$) before impinging on a polarizing beam splitter cube (PBC). The transmitted light (p-polarized after the PBC) is rotated by a second half-wave plate ($\theta_o$) and subsequently sent through a grating pair. A planar silver mirror is placed after the grating pair to end the free-space section and reflect the light back into the fiber loop.

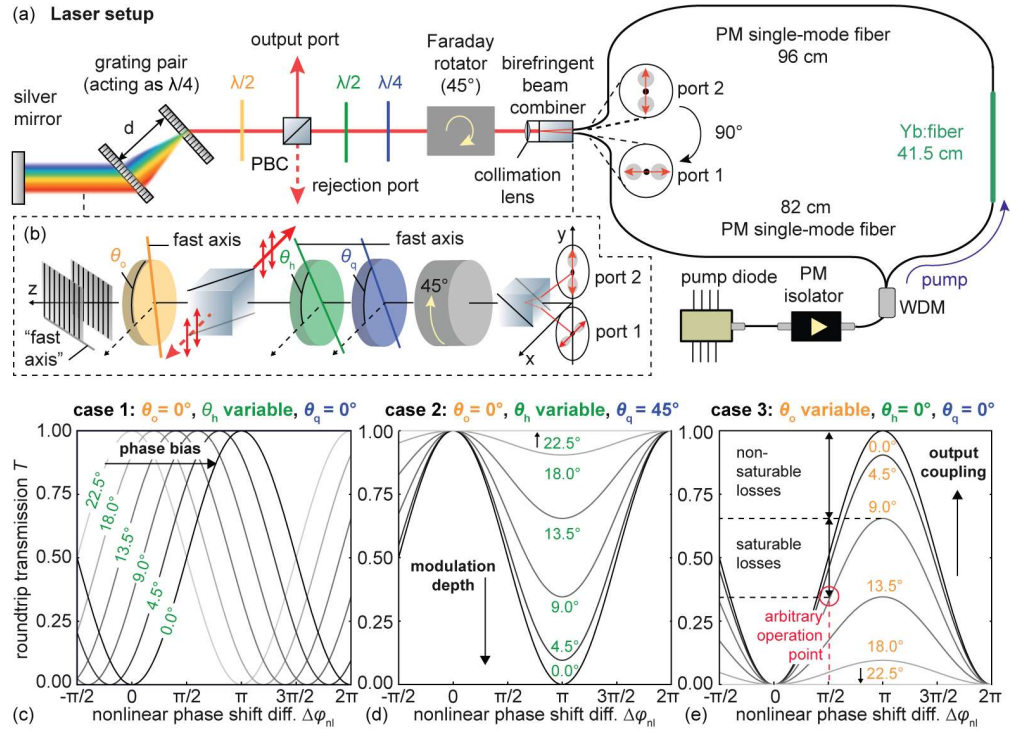

**Fig. 1.** (a) All-PM laser setup featuring a NALM-fiber loop and a free-space section for dispersion control and flexible phase bias/modulation depth/output tuning. (b) Orientation of the polarization-controlling elements in three-dimensional space. (c)-(e) Roundtrip cavity transmission function $\mathcal{T}$ as a function of the nonlinear phase shift difference $\Delta\varphi_{nl} := \varphi_{nl,1} - \varphi_{nl,2}$, where $\varphi_{nl,1}$ is the nonlinear phase accumulated by wave 1 (port 1→ port 2) and $\varphi_{nl,2}$ by wave 2 (port 2→ port 1). Three exemplary wave plate settings are depicted: (c) Case 1 ($\theta_q=0°$) corresponds to a configuration where tuning the angle $\theta_h$ only changes the phase bias, but not the modulation depth. (d) Case 2 ($\theta_q=45°$) represents the other extreme, i.e. a setting where tuning the angle $\theta_h$ only changes the modulation depth, but not the phase bias. (e) Case 3 shows the effect of the second half-wave plate, which acts on the non-saturable/linear losses, i.e. the power obtained at the output port. The power at the rejection port represents the saturable losses, which are minimized in mode-locked operation. By adjusting the three angles $\theta_h$, $\theta_q$ and $\theta_o$ independently, basically any desired combination of phase bias, modulation depth and output coupling ratio can be reached, thus enabling reliable and repeatable mode-locking for a large range of pump powers and dispersion parameters.

## 2.2 Cavity transmission function

In order to understand how the polarization evolution within the free-space section affects the losses and thus the mode-locking behavior, it is useful to take a look at the total cavity roundtrip transmission function $\mathcal{T}$. Total transmission (i.e. $\mathcal{T}=1$) corresponds to a situation where all the light stays within the cavity, while for $\mathcal{T}=0$, the losses are at their maximum and hence no lasing occurs at all. As mentioned in the introduction, self-starting mode-locked operation is obtained when small deviations of the nonlinear phase shift difference $\Delta\varphi_{nl}$ from zero lead to a large decrease in the losses, i.e. when operating on the slope of the cavity transmission $\mathcal{T}$. The roundtrip transmission function $\mathcal{T}$ can be calculated by expressing the effect of each intracavity element in terms of appropriate Jones matrices [40]. The Jones matrices used in our case are listed in table 2.

Table 2. Jones matrices for the intracavity elements

| Component | Jones matrix |
|---|---|
| Half-wave plate | $M_{\lambda/2}(\theta) = e^{-\frac{i\pi}{2}} \begin{pmatrix} \cos^2\theta - \sin^2\theta & 2\cos\theta\sin\theta \\ 2\cos\theta\sin\theta & \sin^2\theta - \cos^2\theta \end{pmatrix}$ |
| Quarter-wave plate | $M_{\lambda/4}(\theta) = e^{-\frac{i\pi}{4}} \begin{pmatrix} \cos^2\theta + i\sin^2\theta & (1-i)\sin\theta\cos\theta \\ (1-i)\sin\theta\cos\theta & \sin^2\theta + i\cos^2\theta \end{pmatrix}$ |
| Faraday rotator | $M_F(\theta) = \begin{pmatrix} \cos\theta & \sin\theta \\ -\sin\theta & \cos\theta \end{pmatrix}$ |
| PBC (transmission/reflection) | $M_{PBC,trans} = \begin{pmatrix} 1 & 0 \\ 0 & 0 \end{pmatrix}, M_{PBC,refl} = \begin{pmatrix} 0 & 0 \\ 0 & 1 \end{pmatrix}$ |
| End mirror | $M_{silver} = \begin{pmatrix} -1 & 0 \\ 0 & -1 \end{pmatrix}$ |
| Fiber loop (i.e. inverting the x-and y-components) | $M_{loop} = \begin{pmatrix} 0 & 1 \\ 1 & 0 \end{pmatrix}$ |
| Nonlinear phase shift | $M_{nl}(\Delta\varphi_{nl}) = \begin{pmatrix} e^{i\Delta\varphi_{nl}} & 0 \\ 0 & 1 \end{pmatrix}$ |

We define the nonlinear phase shift as $\Delta\varphi_{nl} := \varphi_{nl,1} - \varphi_{nl,2}$, where $\varphi_{nl,1}$ is the nonlinear phase accumulated by wave 1 (port 1→ port 2) and $\varphi_{nl,2}$ by wave 2 (port 2→ port 1). Since the gain fiber is placed closer to port 1, coupling the same amount of light into both ports will result in $\Delta\varphi_{nl} > 0$. Note, however, that it is possible to obtain $\Delta\varphi_{nl} < 0$ in case significantly less light is coupled into port 1 than into port 2. The smaller the geometrical asymmetry of the loop, the easier it is to flip the sign of $\Delta\varphi_{nl}$. The asymmetry is not only determined by the length ratio of the single-mode fibers on each side of the gain fiber, but also by the pump direction: the beam co-propagating with the pump will experience stronger amplification at the beginning of the gain fiber and will thus accumulate more nonlinear phase in the rest of the gain fiber than the counter-propagating beam. To calculate the exact "break even" point, i.e. the laser operation point at which the splitting ratio compensates a given geometrical asymmetry, one needs to take into account the gain level/gain saturation behavior, which is beyond the model discussed here. The splitting ratio itself, however, can be directly calculated using the Jones matrix formalism (see section 2.3).

The wave plate angles stated in this paper are given with respect to the x-axis, i.e. the horizontal axis (see Fig. 1 (b)). Furthermore, the birefringent beam combiner element is aligned such that the polarization coupled into the loop at port 1 is parallel to the x-axis, which also corresponds to the polarization direction of the light transmitted through the PBC. The grating pair is oriented such that the light gets dispersed in the x-z-plane.

In addition, we experimentally observed that – unless the incoming polarization was perfectly parallel (s-polarization) or perpendicular to the grating lines (p-polarization) – the light would be elliptically polarized after one pass through the grating pair. The gratings are designed by the manufacturer to provide the same diffraction efficiencies for both s- and p polarization (see Table 1), but no specifications were available with regards to the difference in phase shift for s- and p. Hence, we experimentally determined the polarization state after a single pass through the grating pair by using an auxiliary, linearly polarized laser beam. Power ratio measurements using a PBC after the grating pair confirmed the following: no change in polarization state for s- or p-polarization, a perfectly circularly polarized beam if the polarization impinges on the grating pair at a 45°-angle and elliptical polarization state otherwise. Hence, we modelled the effect of the grating pair on the polarization state by using the Jones matrix of a quarter-wave plate with its fast axis being parallel to the x-axis. The intracavity electric field after roundtrip $\vec{E}_{intra}^{rt}$ can thus be expressed as

$$\vec{E}_{intra}^{rt}\left(\theta_q,\theta_h,\theta_o,\Delta\varphi_{nl}\right) = M_{PBC,trans} M_{\lambda/2}\left(\theta_o\right) M_{\lambda/4\,(gratings)}(0°) M_{silver} M_{\lambda/4\,(gratings)}(0°)$$
$$M_{\lambda/2}\left(\theta_o\right) M_{PBC,trans} M_{\lambda/2}\left(\theta_h\right) M_{\lambda/4}\left(\theta_q\right) M_F\left(45°\right) \quad (1)$$
$$M_{loop} M_{nl}\left(\Delta\varphi_{nl}\right) M_F\left(45°\right) M_{\lambda/4}\left(\theta_q\right) M_{\lambda/2}\left(\theta_h\right) \vec{e}_x$$

where $\vec{e}_x = (1,0)$ is the normalized field vector along the x-axis, i.e. corresponding to the polarization state of the light transmitted by the PBC. The field vectors of the light at the rejection ($\vec{E}_{rej}^{rt}$) and the output port ($\vec{E}_{out}^{rt}$) are given by similar expressions. The cavity roundtrip transmission $\mathcal{T}$ then simply corresponds to

$$\mathcal{T}\left(\theta_q,\theta_h,\theta_o,\Delta\varphi_{nl}\right) = \left|\vec{E}_{intra}^{rt}\right|^2, \quad \text{since} \quad \left|\vec{E}_{intra}^{rt}\right|^2 + \left|\vec{E}_{rej}^{rt}\right|^2 + \left|\vec{E}_{out}^{rt}\right|^2 = 1. \quad (2)$$

In Fig. 1(c)-(e), we show the roundtrip transmission $\mathcal{T}$ as a function of the nonlinear phase shift difference $\Delta\varphi_{nl}$ for three particularly relevant sets of wave plate angles:

- **Case 1: $\theta_o = 0°$, $\theta_h$ = variable, $\theta_q = 0°$:** In this configuration, changing the angle $\theta_h$ of the half-wave plate shifts the transmission function without affecting the modulation depth. In other terms, this configuration allows for continuous tuning of the phase bias only.
- **Case 2: $\theta_o = 0°$, $\theta_h$ = variable, $\theta_q = 45°$:** With the fast axis of the quarter-wave plate being at a 45°-degree angle, the half-wave plate now solely affects the modulation depth without changing the phase bias.
- **Case 3: $\theta_o$ = variable, $\theta_h = 0°$, $\theta_q = 0°$:** At any arbitrary laser operation point, the momentary total losses consist of saturable losses (the rejection port), as well as linear losses (the output port). The saturable losses are automatically minimized when mode-locked operation is achieved. The half-wave plate placed between the PBC and the gratings provides the freedom to adjust the output coupling ratio (i.e. the linear losses) without affecting the phase bias. Furthermore, it also allows for direct "sampling" of the intra-cavity light instead of only having access to the rejected optical spectrum. Note that replacing the output coupling half-wave plate by a quarter-wave plate would cap the linear losses at a maximum of 25% instead of 100% due to the fact that the grating pair already acts as a quarter-wave plate. In practice, the useful modulation range is of course determined by the amount of intracavity power that needs to be sent back into the fiber loop in order to maintain clean mode-locking at a given pump power.

## 2.3 Splitting ratio

In contrast to other implementations of the figure-9 concept that featured a fixed splitting ratio $k$ at the loop entrance [24,25,28,33], the splitting ratio in our case varies as a function of the wave plate positions. The normalized field vector at the entrance of the loop is

$$\vec{E}_{\text{loop entrance}} = M_F(45°) M_{\lambda/4}(\theta_q) M_{\lambda/2}(\theta_h) \vec{e}_x. \quad (3)$$

We define $k$ as the fraction of the light that is coupled into port 1 (polarization along the x-axis), i.e.

$$k = \left| E_{\text{loop entrance, x}} \right|^2. \quad (4)$$

For case 1 mentioned above (i.e. full modulation depth, but varying phase bias) the splitting factor $k$ is 0.5 for all values of the half-wave plate angle $\theta_h$. This 50:50 splitting ratio is due to the fact that after passing through the quarter-wave plate on the way towards the collimator, the light will be elliptically polarized with the major axis of the ellipse being along the x-axis. The Faraday rotator then rotates this ellipse by 45°. Thus, regardless of the ellipticity, the same amount of light will be coupled into both fiber ports whose slow axes are aligned along x and y, respectively. Tuning the quarter-wave plate angle $\theta_q$ between 0° and 45° will vary $k$, while also changing the modulation depth. By tuning the splitting ratio $k$, the loop asymmetry (i.e. the value of $\Delta\varphi_{nl}$) can be influenced without having to change the location of the gain fiber within the loop. Hence, achieving self-starting mode-locked operation depends much less on the exact position of the gain fiber than in laser designs with a fixed splitting ratio.

## 2.4 Operating the laser

The overall laser cavity configuration as depicted in Fig. 1 (a) provides access to a vast parameter space. Most degrees of freedom, such as the pump power, the angular position of the wave plates and the grating separation can be precisely set and are thus easily reproducible. The laser configuration presented here however also has a rather large alignment tolerance, i.e. lasing and even clean mode-locking can be obtained with slightly different angular positions of the end mirror. End mirror alignment can in turn have a significant impact on the center wavelength of the output spectrum: in certain mode-locking configurations, particularly in the negative dispersion regime, we have been able to obtain a center wavelength shift of ~30 nm by slightly tilting the end mirror without losing mode-locked operation. In order to have a reference point for end mirror alignment, we thus have found it most useful to align it at low pump powers such that the lasing threshold is minimized. To find the lowest lasing threshold, it is recommendable to set the quarter-wave plate angle to $\theta_q=0°$, the half-wave plate angle to $\theta_h=22.5°$ and the output coupling wave plate angle such that only a small amount is exiting at the output port (i.e. $T\approx 1$). Once lasing is obtained, the pump power can be ramped up while slowly tuning $\theta_h$. This corresponds to changing the phase bias as depicted in Fig. 1(c) (case 1) and will reliably lead to mode-locked operation. Once the laser is mode-locked, fine-tuning can be done using the other wave plates. Since in practice, the phase shifts induced by the wave plates are wavelength-dependent, this fine-tuning can also be used to slightly change the position/shape of the spectra. Once an initial set of optimal parameters has been found, the wave plates do not need to be adjusted anymore and the laser can be mode-locked only by ramping up the pump power.

When the laser itself serves as the object of scientific interest – as it was the case in our study – damage may occur while trying to explore the limits of the parameter space. The pump current threshold value determined as described above can then be used as a fairly reliable indicator of potential damage. Although lasing may still be achieved at higher pump powers, if the lasing threshold cannot be lowered to the original value, mode-locking will usually not be achieved anymore. Damage can occur during the build-up of Q-switching pulses with uncontrollable peak powers. Such pulses are most likely to occur upon a sudden change of cavity losses at large pump powers; the most extreme example being physically blocking the cavity or misaligning the end mirror to the point where feedback of the light into the fiber loop is momentarily interrupted. In this setup, the wavelength division multiplexer (WDM) was identified as the most vulnerable component. It is worthwhile however to emphasize that, although the fiber loop was re-spliced twice during the course of this study, the different mode-locking states and their performance (including the noise characteristics) were easy to recover even with a freshly spliced fiber loop.

## 3. Dispersion measurements

In this work, we specifically focused on characterizing operation points that lie within different intracavity dispersion regimes. The most straight-forward way to tune the intracavity dispersion in the type of Yb:fiber laser described here is the use of a grating compressor. A rough estimate of the intracavity dispersion can be obtained by summing up the contributions of the individual cavity elements using their respective data sheets. These data, however, often have large tolerances or may only be specified at particular wavelengths. Furthermore, data on the changes in dispersion values as a function of the inversion level in the active gain fiber is usually not available. Here, we explicitly measured the intracavity dispersion for different grating separation settings using a method proposed by Knox [41].

The method relies on inserting a slit in the section of the laser cavity where the optical frequencies are spread out in space, i.e. between the grating compressor and the end mirror (see inset in Fig. 2(a)). Once mode-locking has been achieved, the slit is laterally moved in small steps across the beam, thereby acting as a bandpass filter. At each step, the laser output spectrum is recorded using an optical spectrum analyzer (Ando AQ6315A) (Fig. 2(a)), while the value of the pulse repetition rate $f_{\text{rep}}$ is measured by a photodiode connected to a microwave spectrum analyzer (Keysight PXA N9030B) with its resolution bandwidth set to 3 Hz. To eliminate the influence of small random drifts in repetition rate, the measurement of $f_{\text{rep}}$ was repeated 25 times and averaged before moving to the next slit position. We then extract the wavelength/optical angular frequency $\omega_c$ corresponding to the center of mass of each spectrum and plot the corresponding values of $f_{\text{rep}}(\omega_c)$ as shown in Fig. 2(b). The wavelength-dependent group delay $T(\omega_c)$ values are calculated according to

$$T(\omega_c) = \frac{\partial \phi}{\partial \omega} = \frac{1}{f_{\text{rep}}(\omega_c)} \tag{5}$$

and fitted with a second-order polynomial (solid black lines in Fig. 2(b)). The first derivative of this fit then yields the group delay dispersion (GDD)

$$\text{GDD} = \frac{\partial^2 \phi}{\partial \omega^2} = \frac{\partial T}{\partial \omega}. \tag{6}$$

Note that using a second-order fit for $T(\omega_c)$ corresponds to taking into account second-order (GDD) and third-order (TOD) dispersion, but neglecting higher orders. The obtained GDD-curves are shown as solid lines in Fig. 2(c); the wavelength plotting range corresponds to the range over which mode-locking could be sustained during the measurement. For comparison, the dashed lines indicate the calculated GDD-curves obtained from adding the following contributions: gain fiber and the single-mode fiber (data provided by the vendors), Faraday rotator (20-mm long terbium gallium garnet (TGG) crystal, using Sellmeier equation from [42]) and the grating dispersion calculated according to the standard formula

$$\text{GDD}_{\text{grating}}(\lambda) = -\frac{\lambda^3 d}{\pi c^2 \Lambda_g^2}\left(1 - \left(\frac{\lambda}{\Lambda_g} - \sin(\alpha)\right)^2\right)^{-3/2}, \tag{7}$$

where $d$ denotes the grating separation, $\alpha$ is the angle of incidence and $\Lambda_g = 1$ µm is the grating period.

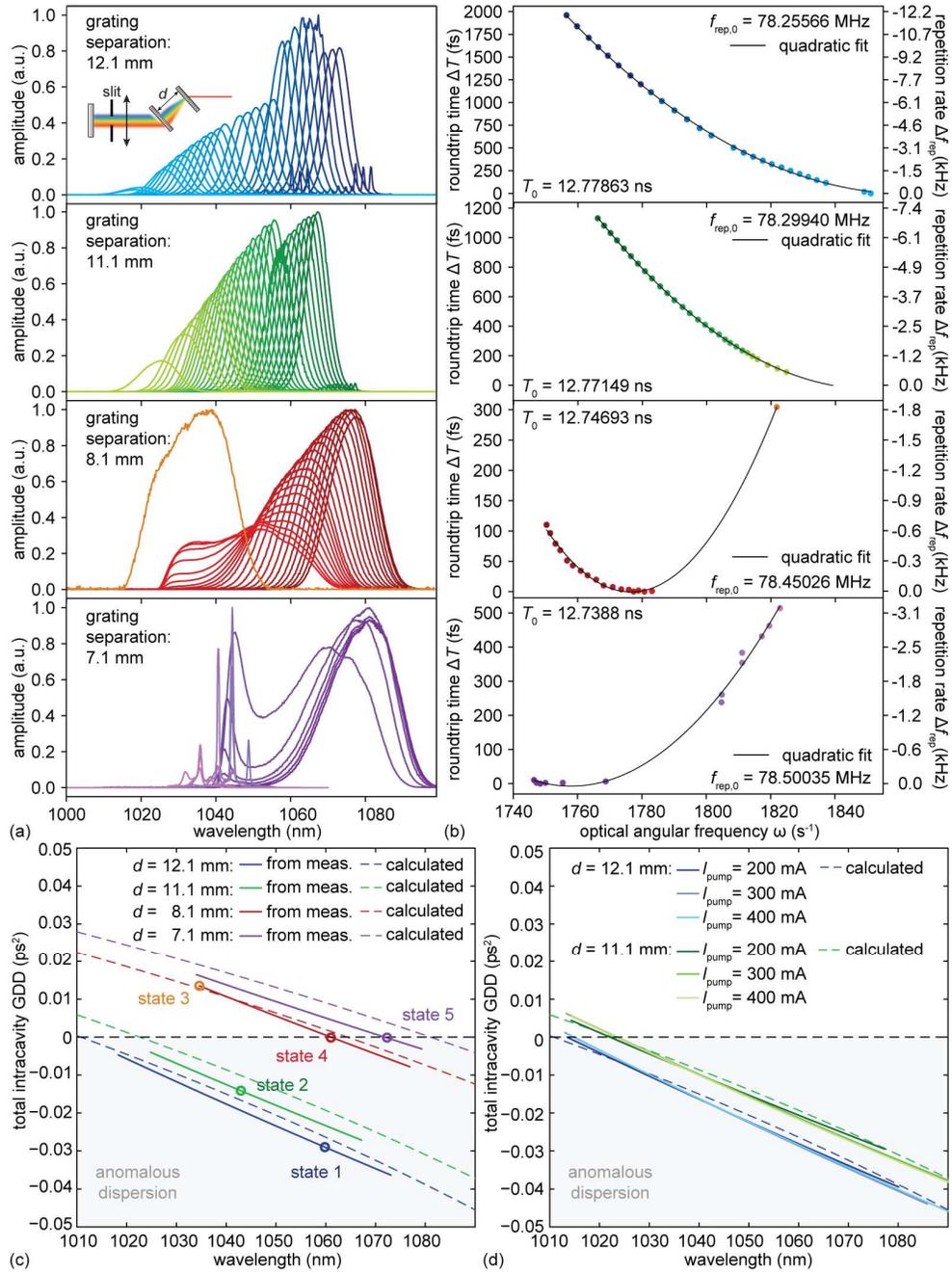

**Fig. 2.** (a) Mode-locked spectra recorded with an optical spectrum analyzer while translating the slit across the beam as shown in the inset. (b) Repetition rate frequency $f_{\rm rep}$ recorded with a microwave spectrum analyzer set to 3 Hz resolution bandwidth (right axis) and corresponding pulse round-trip time/group delay $T=T_0+\Delta T=1/f_{\rm rep}$. The offset values (arbitrarily chosen as the shortest round-trip time/highest repetition rate of the measurement series) have been subtracted for better readability and are stated as annotations in the figure. The black curves are second-order polynomial fits to the group delay values. (c) Solid lines: GDD curves obtained from taking the first derivative of the polynomial fit plotted over the wavelength range where mode-locked spectra were obtained. Dashed lines: calculated dispersion curves. (d) Additional measurement series, where the waveplate positions were kept constant, and only the grating separation was changed.

Figure 2(c) and (d) show two different measurement series: (c) corresponds to the curves obtained using the data shown in (a) and (b). When changing the grating separation in the series shown in (a) and (b), the wave plate positions were changed as well to find the most stable mode-locking states. Furthermore, the measurement series performed at a grating separation of 8.1 mm also includes a state that was found after increasing the pump power and adapting the wave plate positions (orange curve), while keeping the same grating separation. For the data set shown in Fig. 2(d) however, only the grating separation was changed and the measurement was repeated at different pump powers (no changes in wave plate position). These data show that the pump power level does not noticeably change the intracavity GDD. Changing the total losses (and thus the inversion level) by adjusting the wave plate positions on the other hand does have an impact.

## 4. Mode-locking characterization

In the following, we will discuss five characteristic mode-locking states obtained at different values of intracavity dispersion (see states marked in Fig. 2(c)). The optical spectra are depicted in Fig. 3(a), while the microwave spectra of the fundamental repetition rate frequency and its higher harmonics are shown in Fig. 3(b) and (c), respectively.

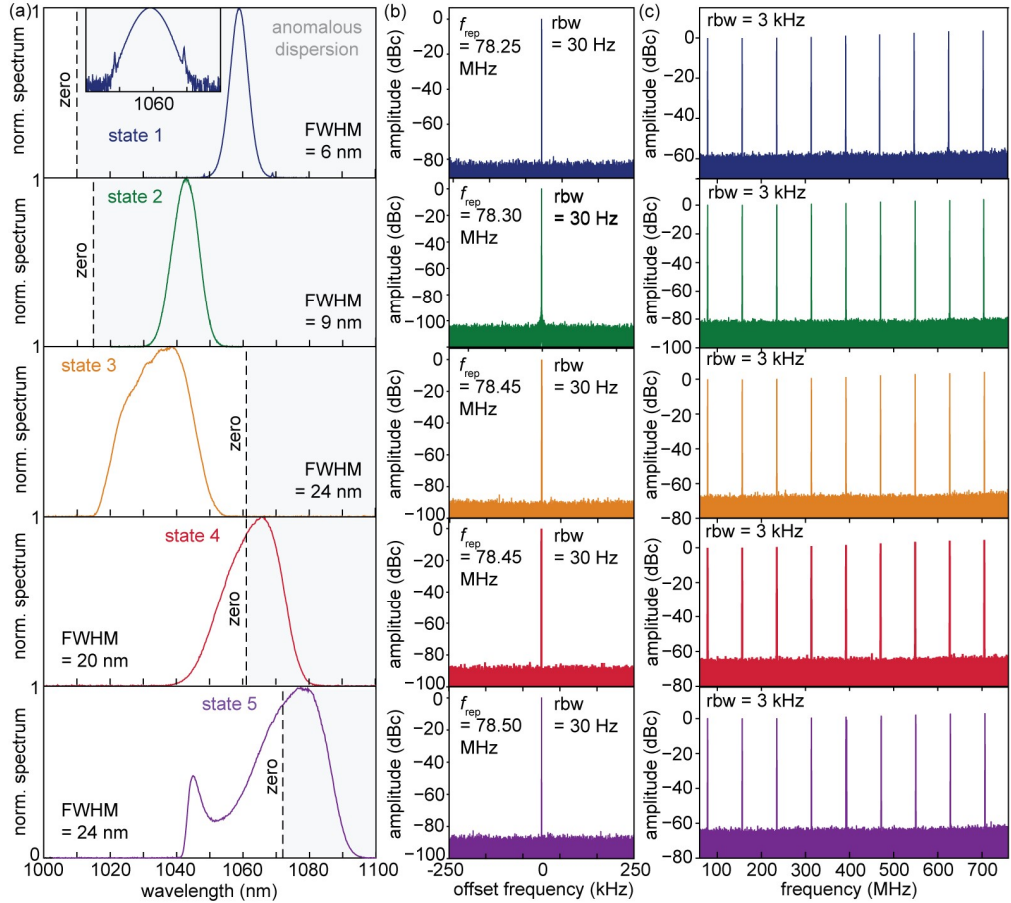

**Fig. 3.** (a) Mode-locking characterization of the 5 states indicated in Fig. 2(c). (a) Optical spectra recorded with an optical spectrum analyzer (Ando AQ6315A) using a resolution bandwidth of 0.5 nm. The gray shaded area indicates anomalous intracavity GDD, while the dashed line represents the wavelength at which the intracavity GDD is zero. (b) 500-kHz zoom into the repetition rate frequency $f_{rep}$ recorded with a microwave spectrum analyzer at a resolution bandwidth of 30 Hz. (c) Wide span showing the higher harmonics of $f_{rep}$.

Table 3. Typical pump power and output power levels for the mode-locking states 1-5

| State | 1 | 2 | 3 | 4 | 5 |
|---|---|---|---|---|---|
| Pump current | 125 mA | 130 mA | 650 mA | 205 mA | 190 mA |
| Pump power | 31 mW | 34 mW | 328 mW | 77 mW | 68 mW |
| Power at output port | 4 mW | 5 mW | 120 mW | 10 mW | 10 mW |
| Power at rejection port | 2 mW | 1 mW | 40 mW | 7 mW | 10 mW |

Note that these mode-locking states were obtained without the slit used during the dispersion measurements. The pump power and waveplate positions for each state were optimized in order to obtain clean fundamental mode-locking (i.e. no multi-pulsing or cw-breakthroughs). Once an initial set of optimal parameters was found, the states could be easily reproduced at will. Table 3 summarizes typical pump power and output power levels for each state (note that the output power in each state is adjustable over a certain range while keeping clean mode-locking by fine-tuning the output coupling wave plate and the pump power level).

All states can be attributed to variations of the stretched-pulse mode-locking regime. In state 1 and 2 (i.e. net negative/anomalous cavity dispersion), solitonic effects limit the intracavity pulse energy, leading to the occurrence of double pulses at higher pump powers. This constraint is lifted for state 3, which is running in the normal dispersion regime around 1030 nm and is capable of delivering much higher output powers (>100 mW). State 4 can be found using the same grating separation as for state 3, however with slightly different wave plate positions, which results in operation around 1060 nm, hence close to zero-dispersion. By slightly reducing the grating separation, the zero-dispersion wavelength can be shifted even further away to ~1075 nm, which is where state 5 operates.

## 5. Amplitude and phase noise measurements

### 5.1 Amplitude noise/relative intensity noise (RIN) measurements

Figure 4(a) shows the amplitude noise/relative intensity noise (RIN) of the five different laser states. Since the output of the pump diode is fiber-coupled and directly spliced to the NALM ring via the WDM for maximum mechanical robustness, the pump power can only be regulated by changing the pump diode's driving current. For the pump diode used here (Thorlabs BLG976-PAG900), we noticed that the RIN is not constant over its operation range (see Appendix for details). We thus included the RIN of the pump measured at the pump current setting corresponding to the different mode-locking states in Fig. 4(a). In Fig. 4(b) we show the RIN of each state normalized by its corresponding pump RIN.

In order to obtain a meaningful data set over a large range of noise frequencies (1 Hz to 1 MHz), we used different measurement methods and combined them to circumvent limitations related to their respective measurement sensitivities. The methods were validated by cross-checking that the data indeed overlaps in regions where different methods were supposed to yield the same results. To minimize the amount of non-laser related noise sources, we omitted the use of external voltage amplifiers and instead used the direct photodiode outputs.

The data traces shown in Fig. 4(a) were obtained as follows:

- For the **RIN characterization of the pump** light (976 nm) we used an InGaAs photodiode (Thorlabs PDA20CS, InGaAs, 5 MHz bandwidth, gain set to 0 dB) whose output was low-pass filtered at 1.9 MHz and recorded with an oscilloscope (LeCroy WavePro 760Zi). 100 time traces of 1 s duration (yielding a resolution bandwidth (rbw) of 1 Hz) and 100 traces of 10 ms duration (rbw=100 Hz) were recorded and directly averaged in the Fourier-domain using the instrument's internal FFT (rectangular window) and averaging function. The final RIN traces were obtained by subtracting the measurement background noise (traces recorded without laser light on the photodiode), normalizing by the corresponding rbw and DC-voltage level and stitching the traces together appropriately.

- The traces for **state 1 and state 2** were obtained by measuring the amplitude noise at the repetition rate frequency (photodiode: Thorlabs DET08 InGaAs bias detector, 5 GHz bandwidth, no amplifier) using a signal source analyzer (SSA, Keysight E5052B). The latter was set to its AM-measurement mode (number of averages: 16), which uses an internal phase-locked loop (PLL) to demodulate the carrier frequency and separate phase noise from amplitude noise.
- For the **states 3, 4 and 5** (which have less noise at low offset frequencies) the SSA's AM-measurement mode did not provide sufficient sensitivity in the range of 1-100 Hz. To cover this range, we instead used the SSA's baseband (BB) noise measurement in combination with the PDA20CS, which has less bandwidth, but a higher voltage output per optical input. At offset frequencies >10 kHz however, the noise level of the internal transimpedance amplifier of the PDA20CS becomes comparable to the RIN of the laser, hence the measurements obtained with the SSA's AM-mode and the DET08 were used to complete the traces at offset frequencies >10 kHz.

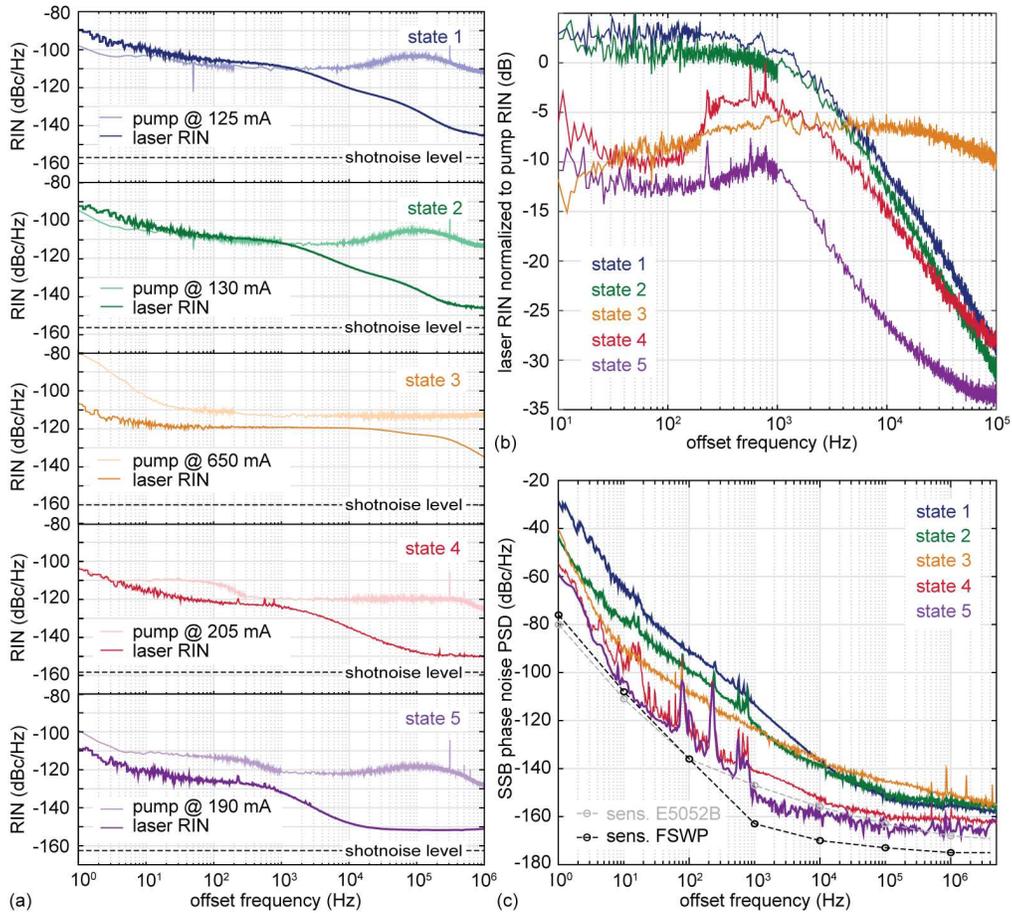

**Fig. 4.** (a) RIN of state 1-5 including the corresponding pump noise. Note that if the trace has been obtained by combining two different measurement methods (see text), the photodetector shot noise level marked here refers to the higher of the two. (b) Laser RIN curves shown in (a) normalized to their respective pump RIN curves. (c) Single-sideband (SSB) phase noise measurement of the repetition rate frequency performed using a signal source analyzer (state 1-4: Keysight E5052B, state 5: Rohde & Schwarz FSWP).

The power spectral densities (PSDs) shown in Fig. 4(a) of both the pump and laser output noise have been integrated over two different offset frequency ranges, yielding the root-mean-square (rms) RIN values listed in Table 4.

Table 4. Integrated rms RIN values for different integration intervals

| State | 1 | 2 | 3 | 4 | 5 |
|---|---|---|---|---|---|
| [1 Hz, 1 MHz] Laser: Pump: | 0.028 % 0.418 % | 0.019 % 0.333 % | 0.048 % 0.230 % | 0.006 % 0.089 % | **0.003 %** 0.081 % |
| [10 Hz, 100 kHz] Laser: Pump: | 0.027 % 0.180 % | 0.018 % 0.141 % | 0.028 % 0.070 % | 0.005 % 0.033 % | **0.002 %** 0.035 % |

*5.2 Discussion of the RIN behavior*

Interesting differences can be observed between the five mode-locking states when looking at the normalized data in Fig. 4(b). In the states operating in the anomalous dispersion regime (state 1 and 2), the laser noise mainly follows the pump RIN level up to ~1 kHz, before rolling off by ~15 dB/decade. For state 4 and 5, we observe a similar roll-off, but in addition also a suppression of the pump noise at frequencies <1 kHz by 5 to 10 dB. State 3 operating in the normal dispersion regime also shows similar pump noise suppression at low offset frequencies, but with a roll-off that starts at much higher offset frequencies (>10 kHz) and has a flatter slope. The fact that the RIN rolls off at lower frequencies for mode-locking states in the negative dispersion regime and close to zero GDD than for states operating with positive intracavity dispersion has already been observed in the case of Yb:fiber lasers based on NPE-mode-locking [43,44]. The noise properties of each state are governed by a complex balance of various effects that need to be taken into account when interpreting these measurements, as will be discussed in the rest of this paragraph.

The response properties of the gain medium alone can be separated from the oscillator dynamics by looking at the noise transfer function in amplifiers made of the same gain medium [45–47]. These references show how the effective roll-off (or corner) frequency of the pump-to-laser noise transfer function not only depends on the fluorescence lifetime of the gain medium (which in the case of Yb would correspond to $f_{\text{roll-off}} \approx 227$ Hz), but also on the absolute pump level. Putting the gain fiber into a resonator generally leads to relaxation oscillations, which in addition to the upper-state lifetime also depend on the resonator losses and length and are often clearly visible as strong peaks in the RIN of cw-lasers. Mode-locking operation however tends to suppress these strong relaxation frequency peaks [48]. Both saturable and inverse saturable absorption have been shown to significantly influence the suppression behavior [49,50]. In a NALM-mode-locked laser, inverse saturable absorption is an inherent effect and simply corresponds to the roll-over of the periodic cavity transmission curve. Low linear losses, i.e. a high cavity finesse is also generally known to support low-noise performance [51].

Depending on the pulse dynamics however, additional effects may come into play. Solitons exchange energy with the co-propagating dispersive wave, leading to Kelly sidebands (in our case, state 1 shows the onset of small Kelly sidebands when looking at the spectrum on a logarithmic scale, see inset in Fig. 3(a)). This exchange stabilizes the pulse energy, but also leads to RIN that is spectrally dependent; it was shown that most of the RIN is actually concentrated in the Kelly sidebands [52]. When the laser is mode-locked entirely in the positive/normal dispersion regime, other self-stabilization mechanisms are dominating. Wavelength filtering has shown to substantially reduce the RIN in these cases and is even necessary to achieve mode-locked operation for larger intra-cavity dispersion values [53]. Although our laser does not contain a fixed band-pass filter, the NALM effect itself is wavelength dependent. Hence, the rejection port can act as a wavelength filter, since the spectral wings of strongly chirped pulses will experience a different nonlinear phase shift in the loop than the pulse center. Another wavelength-dependent noise factor is amplified spontaneous emission (ASE), which is generally stronger around the maximum of the gain profile (i.e. ~1030 in Yb:fibers).

A favorable combination of the effects stated above led to particularly low-noise operation in the case of state 5 (operating around 1075 nm close to zero dispersion). The integrated RIN values of 0.002 % in [10 Hz, 100 kHz] (0.003 % in [1 Hz, 1 MHz]) are among the best values achieved by free-running state-of-the-art lasers [1,24,54].

*5.3 Phase noise/timing jitter measurements*

In addition to the RIN measurements, we also determined the single-side band (SSB) phase noise of the repetition rate frequency for each of the 5 states using a fast photodetector (Thorlabs DET08 InGaAs) and a signal source analyzer (Keysight E5052B for state 1-4 and Rhode & Schwarz FSWP for state 5). The measurement traces are shown in Fig. 4 (c) and the corresponding integrated timing jitter values are stated in Table 5 for different integrations bandwidths.

*5.4 Discussion of the phase noise behavior*

Within the framework of this study, the phase noise/timing jitter measurements were performed with a signal source analyzer as a quick method to compare the different fully free-running mode-locking states. The phase noise values at high offset frequencies and hence the timing jitter values obtained here are however limited by the photodiode + signal source analyzer method and cannot really be compared with the values achieved with state-of-the-art measurements performed using balanced optical cross-correlation (BOC) [25,55,56], optical heterodyning [57] or fiber-delay methods [58,59]. Nevertheless, our measurements already allow us to compare the five different states and observe correlations between the phase and amplitude noise. Free-running phase noise at low offset frequencies (<1kHz) is often said to purely originate from mechanical vibrations and drift of the laser cavity. In our setup however, all five states were obtained from the same laser, i.e. without changing any of its mechanical components nor its environment (the laser was placed on an optical table inside a box covered with a lid to minimize air turbulences). The differences can be explained by the different RIN, which couples to timing jitter via self-steepening and the Kramers-Kronig phase change in the gain medium [60]. A more fundamental limitation comes from ASE, which couples to the timing jitter either directly or via the Gordon-Haus effect [61]. The latter denotes jitter that is caused by fluctuations of the center frequency, which then translates to fluctuations of the group velocity as a consequence of dispersion. It was predicted already in the 1990s that minimizing dispersion in stretched-pulse lasers would lead to a reduction in phase noise/timing jitter [62], which has been experimentally validated by other groups [44,55] and is also consistent with what we observe here.

Table 5. Integrated timing jitter values for different integration intervals (repetition rate $f_{rep}$ = 78 MHz)

| State | 1 | 2 | 3 | 4 | 5 |
|---|---|---|---|---|---|
| [1 Hz, 1 MHz] | 71246 fs | 10119 fs | 11982 fs | 3174 fs | 2166 fs |
| [10 Hz, 1 MHz] | 4014 fs | 1208 fs | 354 fs | 255 fs | 100 fs |
| [100 Hz, 1 MHz] | 803 fs | 351 fs | 202 fs | 81 fs | 72 fs |
| [1 kHz, 1 MHz] | 175 fs | 103 fs | 151 fs | 31 fs | 20 fs |
| [10 kHz, 1 MHz] | 67 fs | 76 fs | 130 fs | 28 fs | 19 fs |

## 6. Carrier-envelope offset (CEO) detection

### 6.1 Self-referencing setup

To complete the characterization of our five mode-locking states, we have detected the CEO frequency using standard *f*-to-2*f* interferometry. In order to reach sufficient peak power for octave-spanning spectral broadening in a photonic crystal fiber (PCF), we amplified the output pulses of the states 1,2,4 and 5 using a 70 cm long PM Yb-doped amplifier pumped/driven by the same type of diode/current controller as the laser. State 3 already provided sufficient output power, hence the amplifier fiber was replaced by a passive PM single-mode fiber in that case in order to keep the rest of the setup as similar as possible. After the amplifier/single-mode fiber, the pulses were compressed using a grating pair (same gratings as used inside the laser, see Table 1) and ~50-60 mW of average power were coupled into a 1-m long commercial highly nonlinear PCF (NKT NL-3.2-945) to generate a coherent octave-spanning supercontinuum. In the *f*-to-2*f* interferometer, light around 1360 nm was frequency-doubled in a 1-mm long periodically poled lithium niobate (PPLN) crystal, before being temporally and spatially overlapped on a photodetector with the 2*f*-component originating from the supercontinuum. The CEO beat notes were recorded using a microwave spectrum analyzer (Keysight PXA N9030B) and are shown in Fig. 5. Note that the amplifier/SCG/*f*-to-2*f* setup was designed for versatility, i.e. to be compatible with all the mode-locking states, and is hence not optimized to provide the best signal-to-noise (SNR) ratio for each state. Nevertheless, the SNR obtained is sufficient for our current main goal: the comparison of the free-running CEO linewidths.

### 6.2 Free-running frequency noise and linewidth analysis

For signals with small phase excursions, the linewidth is sometimes estimated by calculating the integrated phase noise as a function of the lower integration limit: the offset frequency at which the integrated phase noise reaches 1 rad$^2$ corresponds to the half-width at half maximum (HWHM) of the signal [63,64]. Signals exhibiting only small phase excursions can be expressed as a carrier signal with modulation sidebands, and an integrated phase noise of 1 rad$^2$ then corresponds to having ~37 % of the total signal power in the carrier.

However, the free-running CEO signal is subjected to flicker noise (i.e. large drifts at close-in offset frequencies) and hence the small-phase-deviation assumption is not fulfilled. As a consequence, the linewidth has to be stated as a function of measurement time to allow for meaningful comparisons.

Using a sweeping microwave spectrum analyzer with fixed sweep time settings allows us to get a first quick comparison between the five different mode-locking states. In Fig. 5 (b), we show the CEO signal of the five states on a linear scale at 1 kHz rbw in a 10 MHz span. The data was read out after averaging over 100 sweeps, which corresponds to a measurement time of ~8 s. The FWHM linewidths vary by more than two orders of magnitude, ranging from 3 MHz (state 1) down to 9.75 kHz (state 5).

A more quantitatively precise method to determine the linewidth as a function of observation time consists of retrieving the lineshape via the frequency noise PSD $S_f^{CEO}(f)$, whose lowest offset frequency corresponds to the inverse of the measurement time. The lineshape $P_{CEO}(f)$ can be obtained by calculating the Fourier transform of the field autocorrelation function $\Gamma_E(\tau)$, which can be expressed in terms of $S_f^{CEO}(f)$ as described in [65]:

$$P_{CEO}(f) = \mathcal{F}(\Gamma_E(\tau)) = 2\int_{-\infty}^{\infty} e^{-i2\pi f\tau}\Gamma_E(\tau)d\tau = 2\int_{-\infty}^{\infty} e^{-i2\pi f\tau} e^{-2\int_0^{\infty} S_f^{CEO}(f')\frac{\sin^2(\pi f'\tau)}{f'^2}df'} d\tau \quad (8)$$

For a frequency noise PSD that has a constant level, i.e. $S_f^{CEO}(f)=h_0$ (Hz$^2$/Hz), eq. (8) can be analytically solved and yields a Lorentzian profile. A Gaussian profile on the other hand is obtained when modelling the frequency noise as constant below a certain cut-off frequency $f_c$ and zero for offset frequencies $f > f_c$ [65].

In Fig. 5 (b), we added both a Gaussian and a Lorentzian line profile with the same full-width-at-half-maximum (FWHM) to the $f_{CEO}$ measurements of state 1-3. For state 1 and 2, the Gaussian profile is a better fit, while for state 3, the Lorentzian fit is clearly more appropriate. The fact that the line profile of state 3 is closer to a Lorentzian can be traced back to the RIN: the predicted CEO frequency noise PSD $S_f^{CEO, pump}(f)$ caused by pump noise can be expressed according to [66] as

$$S_f^{CEO, pump}(f) = \left(P_{pump} \frac{df_{CEO}}{dP_{pump}}\right)^2 \text{RIN}_{laser}(f) = \xi^2 \cdot \text{RIN}_{laser}(f) \qquad (9)$$

where $P_{pump}$ denotes the pump power and $\xi^2$ is the coupling factor. Since the laser RIN in state 3 exhibits a much higher roll-off frequency (Fig. 4 (b)), the approximation of white frequency noise for $S_f^{CEO}(f)$ is thus more appropriate than for state 1 and 2.

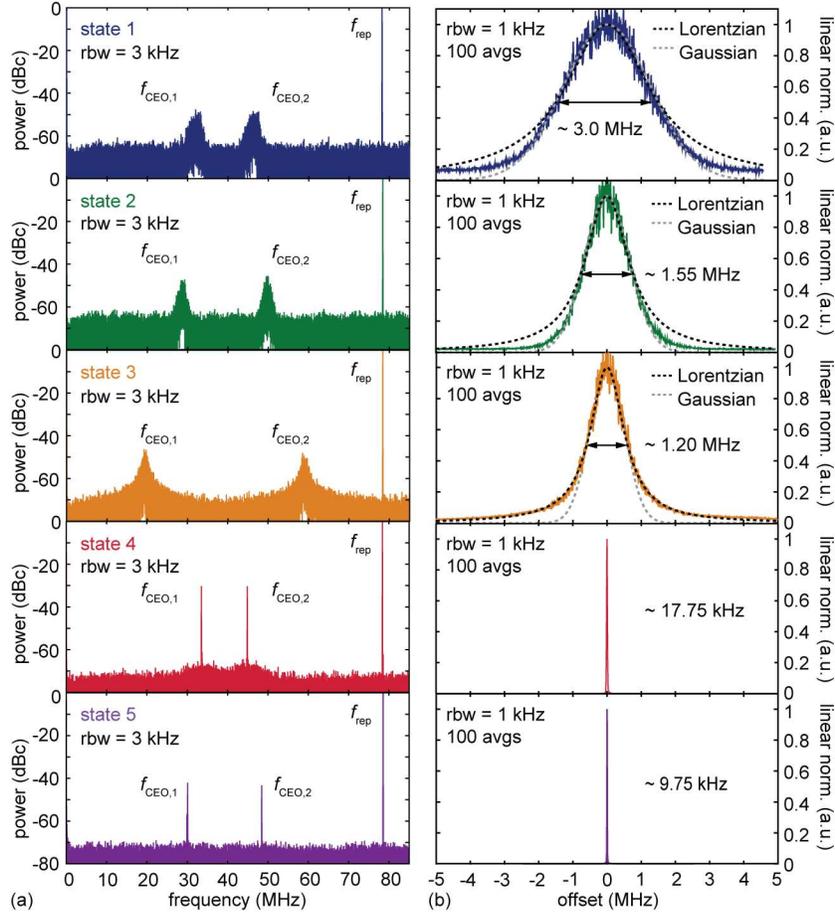

**Fig. 5.** (a) Carrier-envelope offset (CEO) frequency $f_{CEO}$ recorded with a Keysight PXA N9030B microwave spectrum analyzer at a resolution bandwidth (rbw) of 3 kHz. (b) Zoom into $f_{CEO}$ using an rbw of 1 kHz and averaging over 100 sweeps.

In the following, we will take a closer look at the CEO linewidths of state 4 and 5, which are significantly narrower than those of states 1-3. Since both state 4 and 5 operate in a similar pump power range, we set the pump diode current to the exact same value (220 mA) for all the measurements shown in Fig. 6 in order to allow for a straight-forward comparison of the two states. In Fig. 6 (a) and (b), we show a 100-kHz zoom recorded with a 300-Hz resolution bandwidth, corresponding to a total measurement time of ~9 s.

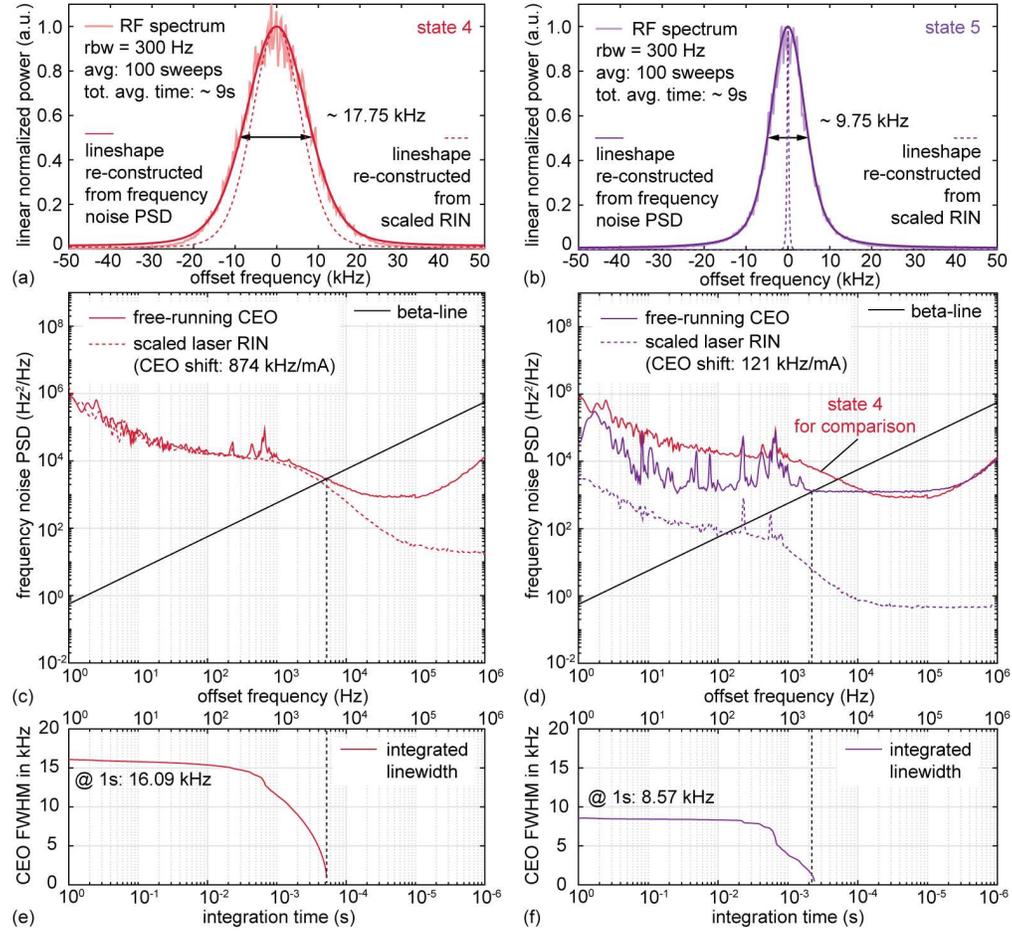

**Fig. 6.** (a),(b) Zoom into $f_{CEO}$ of state 4 and 5 recorded with a Keysight PXA N9030B microwave spectrum analyzer at a resolution bandwidth (rbw) of 300 Hz and averaged over 100 sweeps (total averaging time ~9 s). The solid lines are the lineshapes re-constructed from the frequency noise power spectral densities (PSDs) shown in (c) and (d). (c),(d) Frequency noise PSD $S_f^{CEO}(f)$ measured for state 4 and 5 using the Rohde&Schwarz FSWP signal source analyzer. The "CEO shift" value indicates the shift of $f_{CEO}$ for 1 mA change in pump driver current. The dashed curves are the PSDs $S_f^{CEO,pump}(f)$ corresponding to the contribution of the pump-induced laser RIN to the total frequency noise PSD $S_f^{CEO}(f)$. (e),(f) FWHM of $f_{CEO}$ calculated as a function of integration time using the beta-separation line approach (see text for details).

In Fig. 6 (c) and (d), we show the frequency noise PSD $S_f^{CEO}(f)$ of the free-running CEO of state 4 and 5 measured with a Rohde&Schwarz FSWP SSA. Given this data, we numerically re-constructed the lineshapes according to Eq. (8) using a lower integration limit of 1 Hz (corresponding to 1 s measurement time) and an upper limit of 1 MHz. As can be seen in Fig. 6 (a) and (b), these calculated lineshapes overlap very well with the power spectra obtained by averaging 100 microwave spectrum analyzer sweeps and yield a FWHM of 17.75 kHz (state 4) and 9.75 kHz (state 5).

For comparison, we also calculated the FWHM linewidth as a function of integration time using the simple beta-separation line approximation described in [65,67], which does not require the numerical implementation of Eq. (8). This approach relies on integrating the area $A$ under $S_f^{CEO}(f)$ for $1/T_0 \leq f \leq f_{inter}$, where $T_0$ is the integration time and $f_{inter}$ is the offset frequency at which $S_f^{CEO}(f)$ intersects the beta-line, which is defined as $S_f^{beta}(f)=8\ln(2)f/\pi^2$ (black line Fig. 6 (c) and (d)). The FWHM of the CEO linewidth can then be calculated as

$$\text{FWHM} = \left(8\ln(2)A\right)^{1/2}. \tag{10}$$

For our data, this approach leads to a FWHM of 16.09 kHz (state 4) and 8.75 kHz (state 5) at 1s integration time (Fig. 6 (e) and (f)).

As mentioned above, the pump current/power was the same for the CEO measurements of state 4 and 5. Hence the factor ~2 in linewidth between state 4 and 5 cannot be explained by variations in pump noise, but is a direct consequence of the laser dynamics.

In order to determine the coupling between the laser RIN and the CEO frequency noise PSD as stated in eq. (9), we recorded the shift of $f_{CEO}$ when modulating the pump current around its set value of 220 mA (corresponding to $P_{pump}$= 85.1 mW). For state 4, the shift amounts to 874 kHz/mA (coupling factor $\xi^2 = 172.68\cdot10^{14}$ Hz$^2$) while for state 5, $f_{CEO}$ only shifts by 121 kHz/mA ($\xi^2 = 3.32\cdot10^{14}$ Hz$^2$). We then measured the RIN of the laser output for both states and multiplied the curves by their respective coupling factor $\xi^2$, leading to the dashed curves in Fig. 6 (c) and (d).

For state 4, the scaled RIN curve $S_f^{CEO,pump}(f)$ lies only marginally below the measured PSD $S_f^{CEO}(f)$ at low offset frequencies (at high offset frequencies, the measurement of $S_f^{CEO}(f)$ is limited by the SNR of $f_{CEO}$). This observation suggests that the CEO frequency noise for state 4 is mainly dominated by the laser RIN, which in turn is dominated by pump noise.

For state 5 however, the frequency noise contribution originating from the laser RIN lies significantly below the measured PSD $S_f^{CEO}(f)$. The laser thus operates in a regime where – in contrast to states 1-4 – pump-induced RIN is not the main direct contributor to the $f_{CEO}$ linewidth anymore. The re-constructed theoretical RIN-limited linewidth (dashed line in Fig. 6 (b)) amounts to only 434 Hz. Hence, we suspect the 9.75-kHz-linewidth to be mainly determined by other potential noise sources having an impact on $f_{CEO}$, such as for instance vibrations of the gratings or the end mirror, which could be controlled in a future implementation by using appropriate piezo-actuators.

To the best of our knowledge, the FWHM of 9.75 kHz at 1 s integration time corresponds to the narrowest free-running $f_{CEO}$ linewidth reported to date for NALM-based fiber-lasers.

## 7. Conclusion

In summary, we have presented a versatile PM-NALM-based fiber laser design that offers access to a large parameter space of mode-locking states. The design was implemented in a 78-MHz Yb:fiber configuration using an intracavity grating pair for flexible dispersion control. We discussed the role of the different wave plates in tuning the modulation depth, output coupling ratio and phase bias, thus providing guidelines on how to achieve reliable self-starting mode-locking in this type of lasers. We furthermore presented five representative mode-locking states obtained at different values of intra-cavity dispersion. We characterized the RIN, phase noise and CEO noise in free-running operation to analyze the influence of the mode-locking regime on the overall noise performance of the different states.

We have shown that the coupling of pump noise to laser RIN and thus also to timing jitter and CEO noise can be substantially suppressed for the states operating around zero net intracavity dispersion. In the most extreme case, in addition to being close to zero-dispersion, by operating the laser around 1075 nm center wavelength (i.e. far from the spontaneous emission peak of Yb around 1030 nm), we obtained very low integrated laser RIN of 0.003 % [1Hz, 1 MHz] and a CEO linewidth below 10 kHz. By analyzing the residual contribution of the RIN to the free-running CEO linewidth, we showed that – unlike for most mode-locked lasers – the CEO linewidth in this case is actually not pump-RIN limited anymore.

Various approaches can be employed to actively suppress the RIN in a fiber laser and thus improve the general comb noise performance. However, passive optimization by finding low-noise mode-locking regimes will significantly reduce the requirements on the feedback loop bandwidths necessary for full comb stabilization and hence decrease the system complexity.

Due to its flexibility, robust PM-implementation and compatibility with standard Yb-based amplification schemes, we consider this Yb:fiber laser type to be a promising seed source in particular for low-noise high-power frequency comb applications.

## Appendix A: Pump diode characterization

Upon closer investigation of the pump diode behavior, we observed a correlation between the optical output spectrum and the RIN-level: although the lasing threshold is situated at ~100 mA, the FBG does not fully lock the optical spectrum until the current reaches >400 mA. For pump currents >400 mA the RIN exhibits a fairly constant white noise level for offset frequencies >100 Hz, but rather high low-frequency fluctuations. For pump currents <400 mA (which coincidentally happens to be the range used for most of our mode-locking states), the RIN level is lower in the 1-100 Hz-range, but exhibits a bump around 100 kHz. The overall pump RIN is highest directly above the diode's lasing threshold and seems to reach a minimum just before the proper onset of the FBG-feedback (i.e. pump currents around 200-300 mA). An in-depth analysis of the dynamics within the pump diode causing this type of behavior as well as any generalized statement regarding the applicability of these observations to other pump diodes is however beyond the scope of this paper.

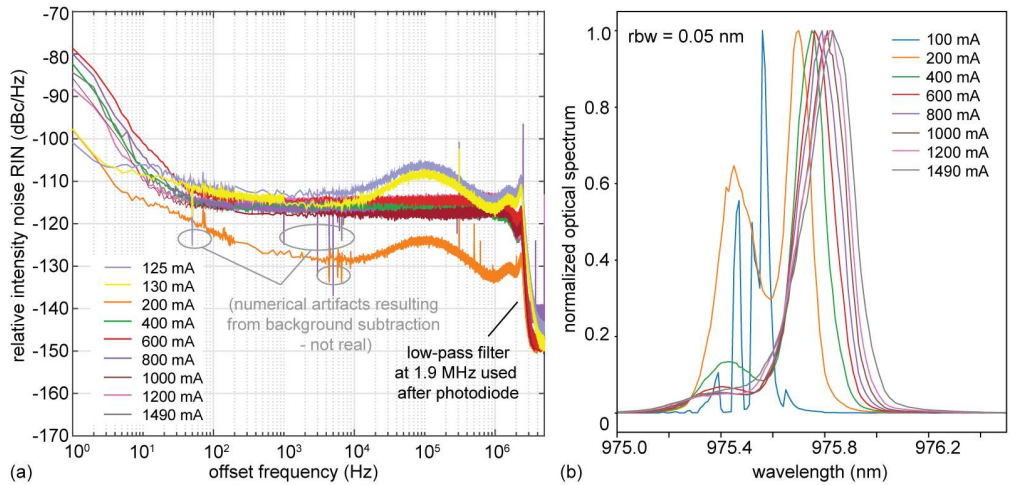

Fig. 7. RIN of the pump diode (Thorlabs BLG976-PAG900) at 25°C for different current values using a CLD1015 controller with its modulation input turned off. (b) Optical spectra of the pump diode output for different driver current values measured with an Ando AQ6315A.

## Funding

Austrian Federal Ministry for Digital and Economic Affairs; National Foundation for Research, Technology and Development; Austrian Science Fund (FWF) (M2561-N36).

## Acknowledgments

The authors would like to thank Valentina Shumakova and Garret Cole for helpful comments on the manuscript.

## Disclosures

The authors declare no conflicts of interest.

# References

1. J. Kim and Y. Song, "Ultralow-noise mode-locked fiber lasers and frequency combs: principles, status, and applications," Adv. Opt. Photonics **8**(3), 465–540 (2016).
2. M. E. Fermann and I. Hartl, "Ultrafast fibre lasers," Nat. Photonics **7**(11), 868–874 (2013).
3. M. Hofer, M. E. Fermann, F. Haberl, M. H. Ober, and A. J. Schmidt, "Mode locking with cross-phase and self-phase modulation," Opt. Lett. **16**(7), 502–504 (1991).
4. V. J. Matsas, T. P. Newson, D. J. Richardson, and D. N. Payne, "Selfstarting passively mode-locked fibre ring soliton laser exploiting nonlinear polarisation rotation," Electron. Lett. **28**(15), 1391–1393 (1992).
5. K. Tamura, H. A. Haus, and E. P. Ippen, "Self-starting additive pulse mode-locked erbium fibre ring laser," Electron. Lett. **28**(24), 2226–2228 (1992).
6. U. Keller, K. J. Weingarten, F. X. Kärtner, D. Kopf, B. Braun, I. D. Jung, R. Fluck, C. Hönninger, N. Matuschek, and J. Aus Der Au, "Semiconductor saturable absorber mirrors (SESAM's) for femtosecond to nanosecond pulse generation in solid-state lasers," IEEE J. Sel. Top. Quantum Electron. **2**(3), 435–451 (1996).
7. O. Okhotnikov, A. Grudinin, and M. Pessa, "Ultra-fast fibre laser systems based on SESAM technology: New horizons and applications," New J. Phys. **6**, 1–22 (2004).
8. D. Popa, Z. Sun, F. Torrisi, T. Hasan, F. Wang, and A. C. Ferrari, "Sub 200 fs pulse generation from a graphene mode-locked fiber laser," Appl. Phys. Lett. **97**(20), 203106 (2010).
9. E. J. Lee, S. Y. Choi, H. Jeong, N. H. Park, W. Yim, M. H. Kim, J. K. Park, S. Son, S. Bae, S. J. Kim, K. Lee, Y. H. Ahn, K. J. Ahn, B. H. Hong, J. Y. Park, F. Rotermund, and D. Il Yeom, "Active control of all-fibre graphene devices with electrical gating," Nat. Commun. **6**, 6851 (2015).
10. Y.-W. Song, S. Yamashita, C. S. Goh, and S. Y. Set, "Carbon nanotube mode lockers with enhanced nonlinearity via evanescent field interaction in D-shaped fibers," Opt. Lett. **32**(2), 148–150 (2007).
11. K. Kieu and M. Mansuripur, "Femtosecond laser pulse generation with a fiber taper embedded in carbon nanotube/polymer composite," Opt. Lett. **32**(15), 2242–2244 (2007).
12. D. Popa, Z. Sun, T. Hasan, W. B. Cho, F. Wang, F. Torrisi, and A. C. Ferrari, "74-fs nanotube-mode-locked fiber laser," Appl. Phys. Lett. **101**(15), 153107 (2012).
13. J. Sotor, G. Sobon, K. Grodecki, and K. M. Abramski, "Mode-locked erbium-doped fiber laser based on evanescent field interaction with $Sb_2Te_3$ topological insulator," Appl. Phys. Lett. **104**(25), 251112 (2014).
14. J. Lee, J. Koo, Y. M. Jhon, and J. H. Lee, "A femtosecond pulse erbium fiber laser incorporating a saturable absorber based on bulk-structured $Bi_2Te_3$ topological insulator," Opt. Express **22**(5), 6165–6173 (2014).
15. N. J. Doran and D. Wood, "Nonlinear-optical loop mirror," Opt. Lett. **13**(1), 56–58 (1988).
16. M. E. Fermann, F. Haberl, M. Hofer, and H. Hochreiter, "Nonlinear amplifying loop mirror," Opt. Lett. **15**(13), 752–754 (1990).
17. I. N. Duling, "Subpicosecond all-fibre erbium laser," Electron. Lett. **27**(6), 544–545 (1991).
18. D. J. Richardson, R. I. Laming, D. N. Payne, M. W. Phillips, and V. J. Matsas, "320 fs soliton generation with passively mode-locked erbium fibre laser," Electron. Lett. **27**(9), 730–732 (1991).
19. J. W. Nicholson and M. Andrejco, "A polarization maintaining, dispersion managed, femtosecond figure-eight fiber laser," Opt. Express **14**(18), 8160–8167 (2006).
20. E. Baumann, F. R. Giorgetta, J. W. Nicholson, W. C. Swann, I. Coddington, and N. R. Newbury, "High-performance, vibration-immune, fiber-laser frequency comb," Opt. Lett. **34**(5), 638–640 (2009).
21. J. Szczepanek, T. M. Kardaś, M. Michalska, C. Radzewicz, and Y. Stepanenko, "Simple all-PM-fiber laser mode-locked with a nonlinear loop mirror," Opt. Lett. **40**(15), 3500–3503 (2015).
22. H. Lin, D. K. Donald, and W. V. Sorin, "Optimizing Polarization States in a Figure-8 Laser Using a Nonreciprocal Phase Shifter," J. Light. Technol. **12**(7), 1121–1128 (1994).
23. T. Jiang, Y. Cui, P. Lu, C. Li, A. Wang, and Z. Zhang, "All PM Fiber Laser Mode Locked with a Compact Phase Biased Amplifier Loop Mirror," IEEE Photonics Technol. Lett. **28**(16), 1786–1789 (2016).
24. D. Kim, D. Kwon, B. Lee, and J. Kim, "Polarization-maintaining nonlinear-amplifying-loop-mirror mode-locked fiber laser based on a 3 × 3 coupler," Opt. Lett. **44**(5), 1068–1071 (2019).
25. N. Kuse, J. Jiang, C.-C. Lee, T. R. Schibli, and M. E. Fermann, "All polarization-maintaining Er fiber-based optical frequency combs with nonlinear amplifying loop mirror," Opt. Express **24**(3), 3095–3102 (2016).
26. W. Hänsel, H. Hoogland, M. Giunta, S. Schmid, T. Steinmetz, R. Doubek, P. Mayer, S. Dobner, C. Cleff, M. Fischer, and R. Holzwarth, "All polarization-maintaining fiber laser architecture for robust femtosecond pulse generation," Appl. Phys. B Lasers Opt. **123**(1), 41 (2017).
27. R. Liao, Y. Song, L. Chai, and M. Hu, "Pulse dynamics manipulation by the phase bias in a nonlinear fiber amplifying loop mirror," Opt. Express **27**(10), 14705–14715 (2019).
28. Z. Guo, Q. Hao, S. Yang, T. Liu, H. Hu, and H. Zeng, "Octave-Spanning Supercontinuum Generation from an NALM Mode-Locked Yb-Fiber Laser System," IEEE Photonics J. **9**(1), 1600507 (2017).
29. G. Liu, X. Jiang, A. Wang, G. Chang, F. Kaertner, and Z. Zhang, "Robust 700 MHz mode-locked Yb:fiber laser with a biased nonlinear amplifying loop mirror," Opt. Express **26**(20), 26003–26008 (2018).
30. W. Liu, H. Shi, J. Cui, C. Xie, Y. Song, C. Wang, and M. Hu, "Single-polarization large-mode-area fiber laser mode-locked with a nonlinear amplifying loop mirror," Opt. Lett. **43**(12), 2848–2851 (2018).

31. M. Lezius, T. Wilken, C. Deutsch, M. Giunta, O. Mandel, A. Thaller, V. Schkolnik, M. Schiemangk, A. Dinkelaker, A. Kohfeldt, A. Wicht, M. Krutzik, A. Peters, O. Hellmig, H. Duncker, K. Sengstock, P. Windpassinger, K. Lampmann, T. Hülsing, T. W. Hänsch, and R. Holzwarth, "Space-borne frequency comb metrology," Optica **3**(12), 1381–1387 (2016).
32. W. Hänsel, M. Giunta, M. Lezius, M. Fischer, and R. Holzwarth, "Electro-optic modulator for rapid control of the carrier-envelope offset frequency," in *2017 Conference on Lasers and Electro-Optics, CLEO 2017 - Proceedings* (Institute of Electrical and Electronics Engineers Inc., 2017), pp. 1–2.
33. Y. Li, N. Kuse, A. Rolland, Y. Stepanenko, C. Radzewicz, and M. E. Fermann, "Low noise, self-referenced all polarization maintaining Ytterbium fiber laser frequency comb," Opt. Express **25**(15), 18017–18023 (2017).
34. S. Salman, Y. Ma, K. Gürel, S. Schilt, C. Li, P. Pfäfflein, C. Mahnke, J. Fellinger, S. Droste, A. S. Mayer, O. H. Heckl, T. Südmeyer, C. M. Heyl, and I. Hartl, "Comparison of two low-noise CEP stabilization methods for an environmentally stable Yb:fiber oscillator," in *Laser Congress 2019 (ASSL, LAC, LS&C)* (The Optical Society of America, 2019), p. JTu3A.17.
35. R. Liao, Y. Song, W. Liu, H. Shi, L. Chai, and M. Hu, "Dual-comb spectroscopy with a single free-running thulium-doped fiber laser," Opt. Express **26**(8), 11046–11054 (2018).
36. R. Li, H. Shi, H. Tian, Y. Li, B. Liu, Y. Song, and M. Hu, "All-polarization-maintaining dual-wavelength mode-locked fiber laser based on Sagnac loop filter," Opt. Express **26**(22), 28302–28311 (2018).
37. J. Fellinger, A. S. Mayer, G. Winkler, W. Grosinger, G.-W. Truong, S. Droste, C. Li, C. M. Heyl, I. Hartl, and O. H. Heckl, "Tunable dual-comb from an all-polarization-maintaining single-cavity dual-color Yb:fiber laser," Opt. Express **27**(20), 28062–28074 (2019).
38. M. C. Stumpf, S. Pekarek, A. E. H. Oehler, T. Südmeyer, J. M. Dudley, and U. Keller, "Self-referencable frequency comb from a 170-fs, 1.5-μm solid-state laser oscillator," Appl. Phys. B **99**(3), 401–408 (2009).
39. T. D. Shoji, W. Xie, K. L. Silverman, A. Feldman, T. Harvey, R. P. Mirin, and T. R. Schibli, "Ultra-low-noise monolithic mode-locked solid-state laser," Optica **3**(9), 995–998 (2016).
40. R. C. Jones, "A New Calculus for the Treatment of Optical Systems I. Description and Discussion of the Calculus," J. Opt. Soc. Am. **31**(7), 488–493 (1941).
41. W. H. Knox, "Dispersion Measurements for Femtosecond-Pulse Generation and Applications," Appl. Phys. B **58**, 225–235 (1994).
42. U. Schlarb and B. Sugg, "Refractive Index of Terbium Gallium Garnet," Phys. status solidi **182**(2), K91–K93 (1994).
43. L. Nugent-Glandorf, T. A. Johnson, Y. Kobayashi, and S. A. Diddams, "Impact of dispersion on amplitude and frequency noise in a Yb-fiber laser comb," Opt. Lett. **36**(9), 1578–1580 (2011).
44. Y. Song, C. Kim, K. Jung, H. Kim, and J. Kim, "Timing jitter optimization of mode-locked Yb-fiber lasers toward the attosecond regime," Opt. Express **19**(15), 14518–14525 (2011).
45. S. Novak and A. Moesle, "Analytic model for gain modulation in EDFAs," J. Light. Technol. **20**(6), 975–985 (2002).
46. H. Tünnermann, J. Neumann, D. Kracht, and P. Weßels, "Gain dynamics and refractive index changes in fiber amplifiers: a frequency domain approach," Opt. Express **20**(12), 13539–13550 (2012).
47. J. Zhao, G. Guiraud, F. Floissat, B. Gouhier, S. Rota-Rodrigo, N. Traynor, and G. Santarelli, "Gain dynamics of clad-pumped Yb-fiber amplifier and intensity noise control," Opt. Express **25**(1), 357–366 (2017).
48. S. Namiki, E. P. Ippen, H. A. Haus, and K. Tamura, "Relaxation oscillation behavior in polarization additive pulse mode-locked fiber ring lasers," Appl. Phys. Lett. **69**(26), 3969–3971 (1996).
49. L. Matos, O. D. Mücke, J. Chen, and F. X. Kärtner, "Carrier-envelope phase dynamics and noise analysis in octave-spanning Ti:sapphire lasers," Opt. Express **14**(6), 2497–2511 (2006).
50. B. R. Washburn, W. C. Swann, and N. R. Newbury, "Response dynamics of the frequency comb output from a femtosecond fiber laser," Opt. Express **13**(26), 10622–10633 (2005).
51. I. L. Budunoğlu, C. Ülgüdür, B. Oktem, and F. Ö. Ilday, "Intensity noise of mode-locked fiber lasers," Opt. Lett. **34**(16), 2516–2518 (2009).
52. C. Wan, T. R. Schibli, P. Li, C. Bevilacqua, A. Ruehl, and I. Hartl, "Intensity noise coupling in soliton fiber oscillators," Opt. Lett. **42**(24), 5266–5269 (2017).
53. P. Qin, Y. Song, H. Kim, J. Shin, D. Kwon, M. Hu, C. Wang, and J. Kim, "Reduction of timing jitter and intensity noise in normal-dispersion passively mode-locked fiber lasers by narrow band-pass filtering," Opt. Express **22**(23), 28276–28283 (2014).
54. J. Chen, J. W. Sickler, E. P. Ippen, and F. X. Kärtner, "High repetition rate, low jitter, low intensity noise, fundamentally mode-locked 167 fs soliton Er-fiber laser," Opt. Lett. **32**(11), 1566–1568 (2007).
55. T. K. Kim, Y. Song, K. Jung, C. Kim, H. Kim, C. H. Nam, and J. Kim, "Sub-100-as timing jitter optical pulse trains from mode-locked Er-fiber lasers," Opt. Lett. **36**(22), 4443–4445 (2011).
56. K. Jung and J. Kim, "Characterization of timing jitter spectra in free-running mode-locked lasers with 340 dB dynamic range over 10 decades of Fourier frequency," Opt. Lett. **40**(3), 316–319 (2015).
57. D. Hou, C.-C. Lee, Z. Yang, and T. R. Schibli, "Timing jitter characterization of mode-locked lasers with <1 zs/√Hz resolution using a simple optical heterodyne technique," Opt. Lett. **40**(13), 2985–2988 (2015).
58. L. A. Jiang, S. T. Wong, M. E. Grein, E. P. Ippen, and H. A. Haus, "Measuring timing jitter with optical cross correlations," IEEE J. Quantum Electron. **38**(8), 1047–1052 (2002).

59. J. Kim, J. Chen, J. Cox, and F. X. Kärtner, "Attosecond-resolution timing jitter characterization of free-running mode-locked lasers," Opt. Lett. **32**(24), 3519–3521 (2007).
60. R. Paschotta, "Noise of mode-locked lasers. Part II: Timing jitter and other fluctuations," Appl. Phys. B **79**, 163–173 (2004).
61. J. P. Gordon and H. A. Haus, "Random walk of coherently amplified solitons in optical fiber transmission," Opt. Lett. **11**, 665–667 (1986).
62. S. Namiki and H. A. Haus, "Noise of the stretched pulse fiber laser: Part I - Theory," IEEE J. Quantum Electron. **33**(5), 649–659 (1997).
63. D. G. Matei, T. Legero, S. Häfner, C. Grebing, R. Weyrich, W. Zhang, L. Sonderhouse, J. M. Robinson, J. Ye, F. Riehle, and U. Sterr, "1.5 μ m Lasers with Sub-10 mHz Linewidth," Phys. Rev. Lett. **118**(26), 263202 (2017).
64. J. L. Hall and M. Zhu, "An introduction to phase-stable optical sources," in *Proceedings of the Internationl School of Physics "Enrico Fermi" Course 118* (1992), pp. 671–702.
65. G. Di Domenico, S. Schilt, and P. Thomann, "Simple approach to the relation between laser frequency noise and laser line shape," Appl. Opt. **49**(25), 4801–4807 (2010).
66. J. J. McFerran, W. C. Swann, B. R. Washburn, and N. R. Newbury, "Suppression of pump-induced frequency noise in fiber-laser frequency combs leading to sub-radian $f_{ceo}$ phase excursions," Appl. Phys. B **86**(2), 219–227 (2007).
67. N. Bucalovic, V. Dolgovskiy, C. Schori, P. Thomann, G. Di Domenico, and S. Schilt, "Experimental validation of a simple approximation to determine the linewidth of a laser from its frequency noise spectrum," Appl. Opt. **51**(20), 4582–4588 (2012).